\documentclass{iau}
\usepackage{graphicx} % Required for inserting images
\usepackage{amsmath}
\usepackage{multirow}
\usepackage{subfigure}
\usepackage[version=4]{mhchem}

\title{IAU384_Siebert_Proceeding}
\author{Mark Siebert}
\date{November 2023}

\begin{document}

\lefttitle{Mark A.\ Siebert \& Anthony J. Remijan}
\righttitle{Clues to PPN Chemical Evolution: The Unique Molecular Environment of V510 Pup}

\jnlPage{1}{7}
\jnlDoiYr{2021}
\doival{10.1017/xxxxx}

\aopheadtitle{Proceedings IAU Symposium}
\editors{C. Sterken,  J. Hearnshaw \&  D. Valls-Gabaud, eds.}

\title{Clues to PPN Chemical Evolution: The Unique Molecular Environment of V510 Pup}

\author{Mark A. Siebert$^1$ \& Anthony J. Remijan$^2$}

\affiliation{$^1$Department of Astronomy, University of Virginia, Charlottesville, VA 22904, USA \\ email: {\email{mas5fb@virginia.edu}} \\
$^2$National Radio Astronomy Observatory, Charlottesville, VA 22903, USA}

\begin{abstract}
V510 Pup (IRAS 08005-2356) is a binary post-AGB system with a fast molecular outflow that has been noted for its puzzling mixture of carbon- and oxygen-rich features in the optical and infrared. To explore this chemical dichotomy and relate it to the kinematics of the source, we present an ACA spectral line survey detailing fourteen newly detected molecules in this pre-planetary nebula. The simultaneous presence of CN/\ce{C2H}/\ce{HC3N} and \ce{SO}/\ce{SO2} support the previous conclusion of mixed chemistry, and their line profiles indicate that the C- and O-rich material trace distinct velocity structures in the outflow. This evidence suggests that V510 Pup could harbor a dense O-rich central waist from an earlier stage of evolution, which persisted after a fast C-rich molecular outflow formed. By studying the gas phase composition of this unique source, we aim to reveal new insights into the interplay between dynamics and chemistry in rapidly evolving post-AGB systems.
\end{abstract}


\maketitle
\section{Introduction}
The nebulae surrounding evolved stars undergoing mass loss represent important sites of recycling for metal-rich gas and dust in our Galaxy. As low--intermediate mass stars leave the Asymptotic Giant Branch (AGB), their circumstellar envelopes undergo dramatic changes: transitioning from slow, spherical winds to complex bipolar/multipolar outflows seen in planetary nebulae (PNe) \citep{Sahai2011}. These dynamics are often met with a chemical transition as well, with molecules and dust being increasingly affected by conditions including high-velocity shocks, internal UV radiation, and AGB dredge up events \citep{Pardo2007,Weiss2009}. The mechanisms for the physical and chemical changes during the transitional pre-planetary nebula (PPN) phase are still not fully understood, but recent work has shown that binary and common envelope interactions are key to this evolution \citep[e.g.~][]{Sahai2018}. 

One puzzling class of object that is critical to understanding the chemical output of evolved stars are PPNe showing signs of mixed chemistry. Stellar outflows are designated as either carbon- or oxygen-rich depending on the C/O ratio of the central star, which typically determines the molecular and dust grain composition throughout the wind \citep{Hofner2018}. However, it appears that this dichotomy is not universal, especially as stars evolve off the AGB. Evidence of this includes the widespread IR observations of fullerene and PAH features toward O-rich PNe \citep{Gielen2011}, as well as the existence of “silicate carbon stars” which have C-rich photospheres but are surrounded by O-rich circumstellar dust, and often show OH masers \citep{Ohnaka2013}. 

V510 Pup (IRAS 08005-2356) is post-AGB star surrounded by a young high-velocity outflow that shows clear evidence of mixed chemistry. It exhibits OH maser emission typical of O-rich evolved stars \citep{Zijlstra2001}, but also shows optical absorption lines from \ce{C2} and \ce{CN}  \citep{Bakker1997}. The extreme dynamics of V510 Pup were characterized recently by \citet{Sahai2015}, where broad (${\sim}$200 km/s) line profiles were observed in CO and SiO using the Submillimeter Array (SMA). Analysis of the velocity gradient across the nebula showed that radiation pressure alone is insufficient to drive the outflow. Optical observations later discovered that V510 Pup is part of a radial-velocity binary, and that its H$\alpha$ line profile modulates with the seven-year orbital period, confirming that the high-velocity bipolar outflow originates from accretion onto a close binary companion \citep{Manick2021}. This makes it the only extreme PPN whose jet origin is directly observed.

V510 Pup is at a critical point in its evolution to offer new insights into chemical transitions and binary interaction in exotic PPNe, but it lacks key constraints on its gas phase chemical inventory. To address this, a detailed spectroscopic study of this source is needed to determine the molecular abundances and their relationship to kinematic structures within the wind. We used interferometric observations from the Atacama Compact Array (ACA) to perform these measurements, and reveal the unique dynamics of C- and O-rich material in this source. %In Section \ref{section:obs} we describe our data set and reduction procedure, in Section \ref{section:results} we present the spectral analysis, and in Section \ref{section:conclusions} we discuss our new picture of this source and its general implications on mixed chemistry in circumstellar winds.

\section{Observations}
\label{section:obs}
We utilize archival data from a Cycle 6 ACA filler program which observed 16 evolved stars transitioning between the AGB and PNe phases (Project code 2018.A.00047.S). These data include spectral line observations of V510 Pup covering the entirety of Bands 3 and 6 (85--115\,GHz, and 211--274\,GHz) at $1{\sim}3$\,km/s velocity resolution. Calibrated visibilities were produced using the standard pipeline of the Common Astronomy Software Application \citep[CASA;][]{McMullin2007}, and each spectral window was then continuum-subtracted, then imaged and deconvolved using the \texttt{tclean} CASA task with briggs weighting. The resulting rms sensitivity ranges between 5 and 13\,mJy/beam among the 56 spectral windows. Some spectral windows overlap in their frequency coverage, so these data were combined in the visibility plane to improve sensitivity. The ACA beam size is $10.1"$--$11.5"$ and $4.3"$--$5.3"$ at Bands 3 and 6, respectively. This is notably larger than the ${\sim}2"$ nebular extent of V510 Pup observed by \citet{Sahai2015}, meaning a spatial analysis was not possible with these data, and our study is purely spectroscopic.

\section{Results \& Analysis}
\label{section:results}
\subsection{Detected species and line profiles}
\label{section:profs}
We detect emission from a total of 16 molecules and isotopologues toward V510 Pup, 13 of which have not yet been observed in this source. A list of all detected species is shown in Table \ref{tab:mols}, including the number of lines observed of each, and whether they are typically found in C- or O-rich sources, using \citet{Ziurys2006} as reference. All these molecules are common in evolved stars environments, including ``parent" molecules originating from the stellar atmosphere (e.g.\ \ce{SO}, \ce{HCN}, \ce{CS}), as well as species produced through photochemistry like \ce{HCO+}, \ce{HNC}, \ce{HC3N}, and \ce{C2H} \citep{Woods2003}. The remarkable result seen in Table \ref{tab:mols} is the presence of characteristic O-rich molecules (\ce{SO}, \ce{SO2}) paired with C-rich species (\ce{HNC}, \ce{CN}, \ce{C2H}, and \ce{HC3N}). This is the first AGB-related source with simultaneous detections of all these molecules, lending support to previous classifications of mixed chemistry.

\begin{table}[t!]
 \centering
 \caption{Summary of detected molecules and isotopologues}\label{tab:mols}
 {\tablefont\begin{tabular}{ccccc}
    \midrule
    Molecule & No.\ of transitions & Range in E$_{\mathrm{up}}$ (K) $^{a}$ &
     Velocity Component(s)$^{b}$&
     Chemistry$^{c}$\\
    \midrule
    \ce{^{12}CO}& 2 & 10 -- 30 & CV, MV, HV &  Both\\
    \ce{^{13}CO} & 2 & 10 -- 30 & CV, MV, HV &  Both\\[1mm]

    \ce{SiO}& 3 &   6 -- 45 &  CV, MV, HV& Both\\
    \ce{^{29}SiO}& 2 &   31 -- 45&CV, MV, HV& Both\\
    \ce{^{30}SiO}& 2 &   31 -- 45&CV, MV, HV& Both\\[1mm]
    \ce{HCN}& 1 &   26 &  CV, MV & Both\\
    \ce{H^{13}CN}& 1 &   25 &CV, MV & Both\\[1mm]
    \ce{CS}& 1 &   35 & CV(?), MV & C-rich\\
    \ce{C^{34}S}& 1 &   28 & MV & C-rich\\[1mm]
    \ce{HCO+}& 1 &   26 & CV, MV & Both\\
    \ce{SO}& 5 & 44 -- 56 & CV & O-rich\\
    \ce{SO2}& 9 & 24 -- 94 & CV & O-rich\\
    \ce{HNC}& 1 &   26 & MV & C-rich\\
    \ce{CN}& 1 &  16 & MV & C-rich\\
    \ce{C2H}& 1 &   25 & MV & C-rich\\
    \ce{HC3N}& 7 & 131 -- 203 & MV & C-rich\\
    \midrule
    \end{tabular}}
\tabnote{\textit{Notes}: $^{a}$Upper state energies obtained from Cologne Database of Molecular Spectroscopy (CDMS) M\"uller et al.\ 2005. $^{b}$LSR velocity ranges of the CV, MV, and HV components are defined in Section \ref{section:profs} and shown in Fig.\ \ref{fig:line_profs}. $^{c}$Denotes whether a particular molecule is typically associated with a carbon- or oxygen-rich surface chemistry in circumstellar envelopes, using the list from \citet{Ziurys2006}.}
\end{table}

\begin{figure}[t!]
\begin{center}
{\includegraphics[width=\linewidth]{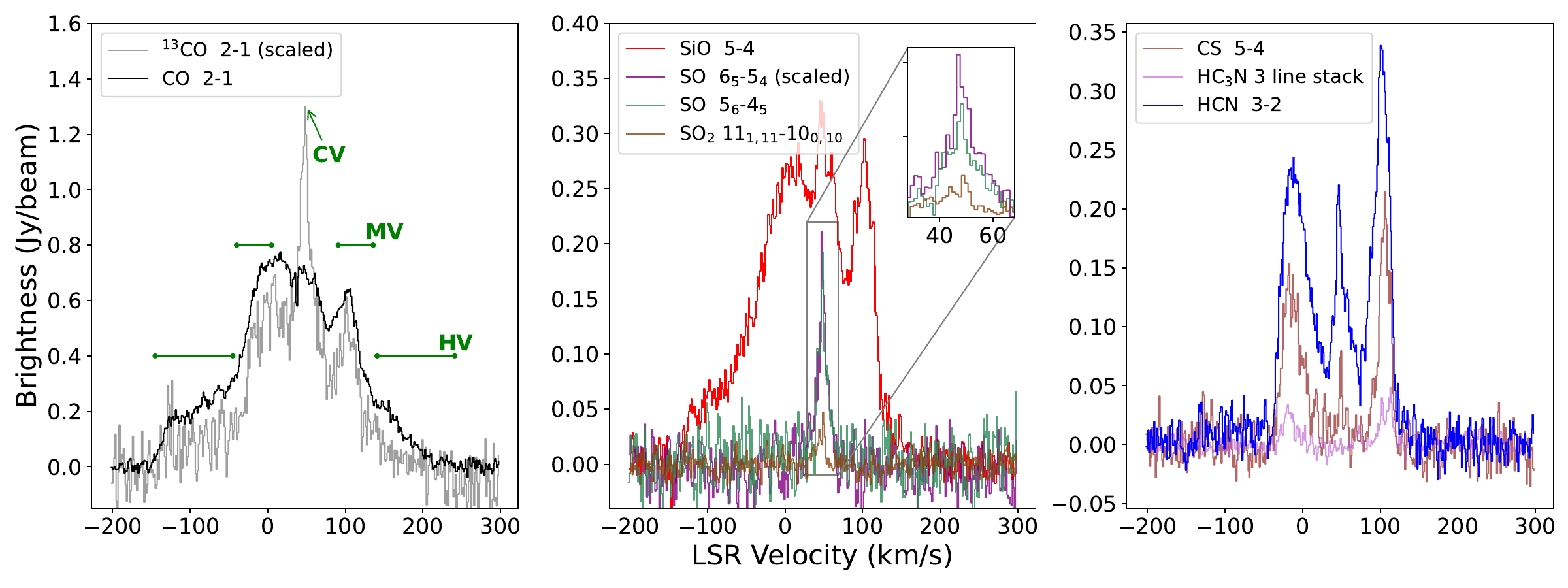}}
\caption{Emission line profiles of various molecules measured by the ACA toward V510 Pup. The central velocity (CV), medium-velocity (MV), and high-velocity (HV) components are defined in the left panel. Species that are typically associated with O- and C-rich evolved stellar outflows are shown in the middle and right panels, respectively. Spectra were extracted from an aperture with diameter 10" centered on the position of the post-AGB star. Some transitions are scaled for comparison.}
\label{fig:line_profs}
\end{center}
\end{figure}

The peculiar chemistry of V510 Pup can be further explored using the kinematic information provided by molecular line profiles. Figure \ref{fig:line_profs} shows a sample of these spectra. The full extent of the extreme outflow can be seen in the broad \ce{CO} and \ce{SiO} line profiles which span the range between $-150$ and $200$\,km/s. Following \citet{Sahai2015}, we decompose this into three main velocity components: a central component (CV) centered on the system velocity ($V_{\mathrm{sys}}=48$\,km/s), a intermediate component (MV) at $V_{\mathrm{sys}}\pm50$\,km/s, and high-velocity wings (HV) at $V_{\mathrm{sys}}\pm110$\,km/s, all shown in Fig.\ \ref{fig:line_profs}. With the improved sensitivity on the \ce{^{13}CO} $J=2-1$ line now available, we note that the CV component has a narrow feature (width $<10$\,km/s) appearing on top of the broad line profile, similar to how this transition was appears toward the similar extreme PPN HD 101584 \citep{Olofsson2019}. This narrow CV component is also observed isolated from the MV and HV components in the emission profiles of \ce{SO}, \ce{SO2}, and \ce{HCN}. Table \ref{tab:mols} also lists which velocity components are detected for each molecule. We find that all molecules which are primarily associated with carbon-rich outflows appear solely in the MV component, while oxygen-rich species are located in the narrow CV region. This can be seen in the middle and right panels of Fig.\ \ref{fig:line_profs}.

\begin{figure}[t!]
\begin{center}
{\includegraphics[width=0.6\linewidth]{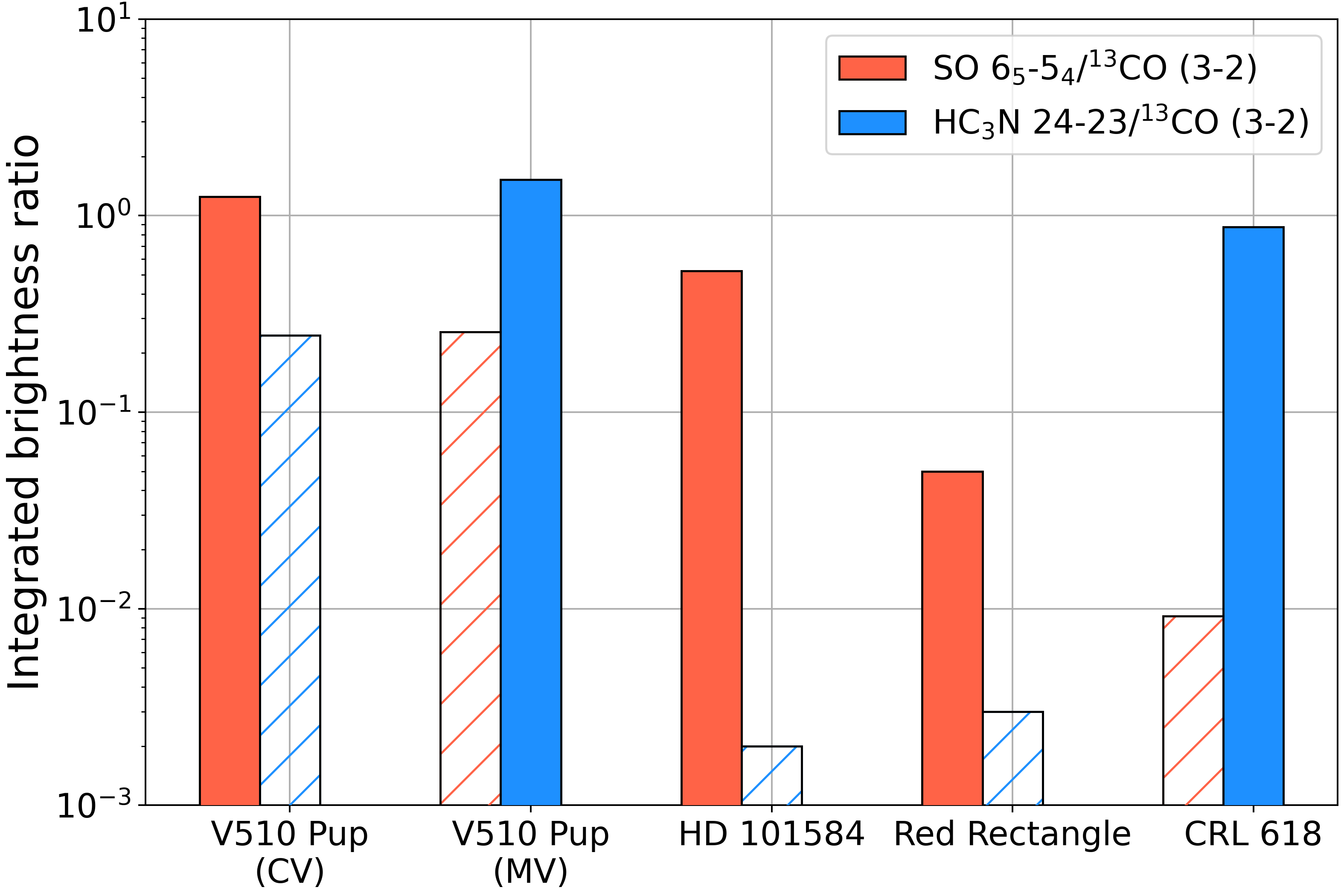}}
\caption{Comparison of integrated molecular line brightness (relative to \ce{^{13}CO})  for \ce{SO} (red) and \ce{HC3N} (blue) among a sample of nearby PPNe. For V510 Pup, the O-rich central and C-rich medium velocity components are labeled CV and MV, respectively. Hatched regions denote upper limits, while solid regions indicate detected transitions. Data for sources other than V510 Pup adapted from \citet{Olofsson2019}, \citet{GallardoCava2022}, and \citet{Pardo2004}.}
\label{fig:ratios}
\end{center}
\end{figure}

Figure \ref{fig:ratios} shows the measured brightnesses of \ce{SO} ($6_5$--$5_4$) and \ce{HC3N} (24--23) relative to \ce{^{13}CO} compared with more well-studied PPNe for which these transitions have been observed previously \citep{Olofsson2019,Pardo2004}. From this, we see that the CV component of V510 Pup is chemically similar to O-rich sources like HD 101584, while the MV component is like the prototypical C-rich PPN CRL 618. Furthermore, this comparison provides evidence that the narrow CV is not depleted in its molecular chemistry as is the case for the Red Rectangle and other ``disk-prominent" post-AGBs \citep{GallardoCava2022}.

\subsection{LTE modeling}

To obtain estimates on the abundances of detected molecules, we model the observed transitions assuming local thermodynamic equilibrium (LTE). This assumption is necessary since we lack spatial data and constraints on the physical conditions of the velocity components to perform a full radiative transfer analysis. To simulate line profiles and fit the transitions, we use the publicly available package \texttt{molsim} \citep{Lee2021}. For each velocity component, we perform a least-squares minimization of the column density, rotational temperature, and line width of a given molecule. For species where only one transition was available, we fix the rotational temperature to that obtained for SO (CV) or \ce{HC3N} (MV). We employ the same procedure for \ce{^{13}CO}, simultaneously fitting the overlapping CV, MV, and HV components. Then, assuming a \ce{^{13}CO}/\ce{H2} ratio of $3.9\times10^{-5}$ \citep{Sahai2015}, we use the relative column densities to calculate final abundances. While this eliminates the need to adopt an explicit angular source size, it does assume that this size is equal to that of \ce{^{13}CO} emission (whether in the CV or MV component). 

\begin{figure*}
\label{fig:simul}
\centering     %%% not \center
\subfigure{\label{fig:a}\includegraphics[width=0.48\linewidth]{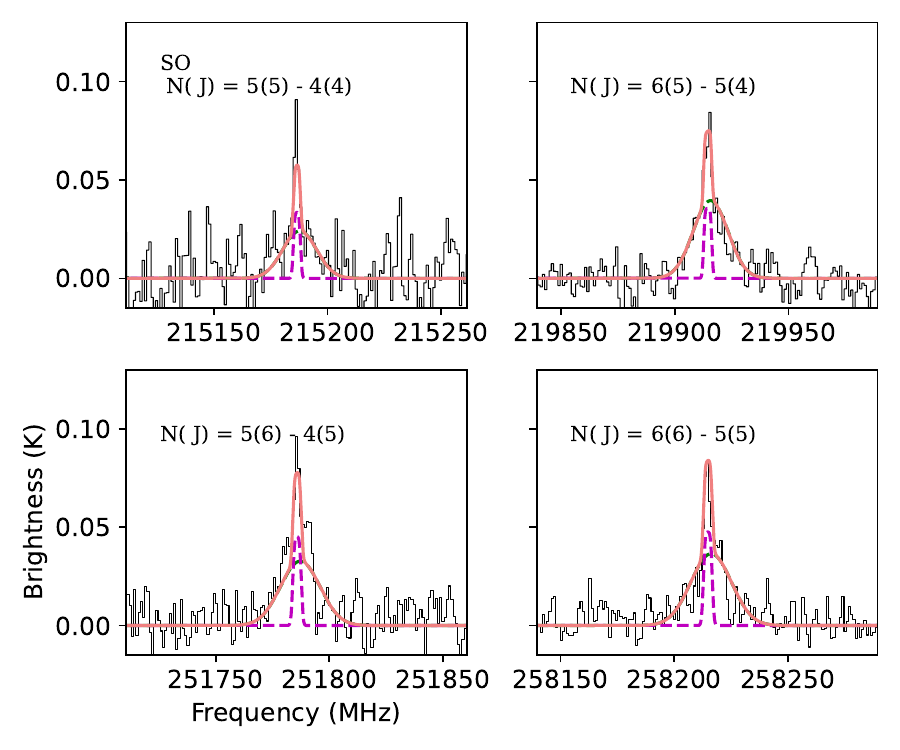}}
\subfigure{\label{fig:b}\includegraphics[width=0.48\linewidth]{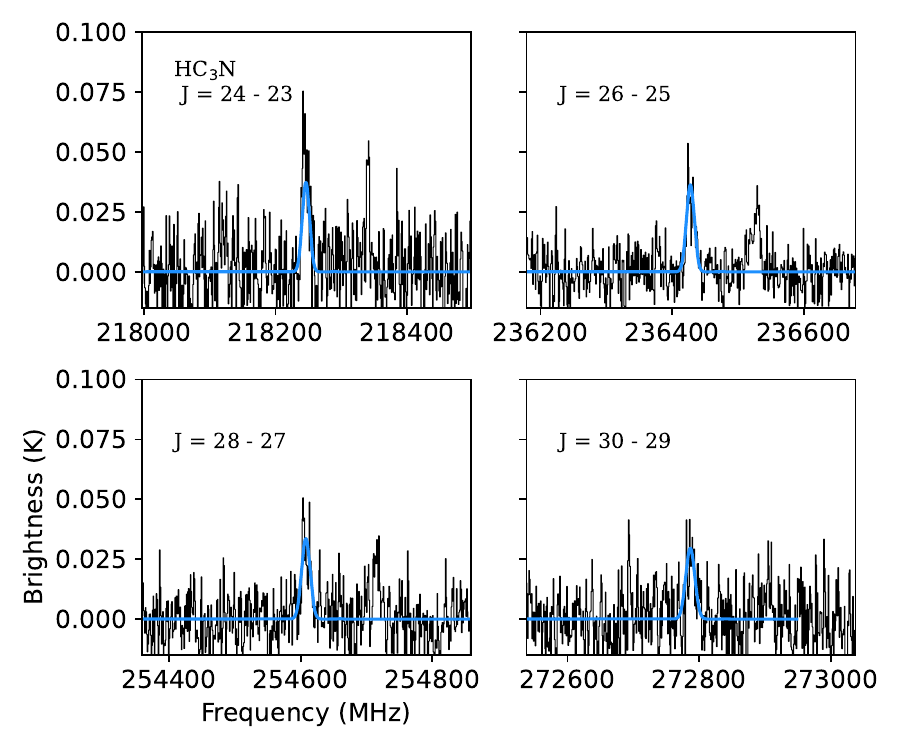}}
\caption{Examples of LTE emission line fits for \ce{SO} (left panels) and \ce{HC3N} (right panels). ACA spectrum shown in black, while colored lines represent fitted \texttt{molsim} simulation. For \ce{SO}, a narrow (magenta dashed) and wide (green dashed) gaussian were needed to reproduce the shape of the CV component. For \ce{HC3N}, one component was used, centered on the red-shifted side of the MV region.}
\end{figure*}

\begin{table}[h!]
 \centering
 \caption{Measurements on physical conditions of optically thin species in the outflow of V510 Pup}\label{tab:LTE}
 {\tablefont\begin{tabular}{cccc}
    \midrule
    Molecule & Component & T$_{\mathrm{rot}}$ (K) &
     $f_{\mathrm{H2}}$ (err.)\\
    \midrule
    \ce{SO}& CV & $43\pm13$ & $4.5\times10^{-6}$ $(1.0)$ \\
    \ce{SO2}& CV & $57\pm18$ & $4.8\times10^{-6}$ $(0.9)$ \\
    \ce{HC3N}& MV & $63\pm16$ & $1.6\times10^{-6}$ $(0.6)$ \\
    \ce{CS}& MV &   63$^{a}$ &  $5.1\times10^{-6}$ $(0.5)$ \\
    \ce{CN}& MV &  63$^{a}$ & $1.3\times10^{-5}$ $(0.1)$ \\
    \ce{C2H}& MV &   63$^{a}$ & $2.8\times10^{-6}$ $(0.3)$ \\
    \midrule
    \end{tabular}}
\tabnote{\textit{Notes}: $^{a}$Rotational temperature of \ce{HC3N} adopted for abundance calculation.}
\end{table}

The results of this analysis are shown in Figure \ref{fig:simul} and Table \ref{tab:LTE}. We find that the CV component exhibits excitation temperatures between 40--60\,K, while the MV component (measured by \ce{HC3N}) is about 63\,K. These values are in line with the temperatures found in the slowly expanding equatorial waist and the bipolar outflow of HD 101584 \citep{Olofsson2019}. The abundances of molecules observed in the MV component are close (within a factor of five) to their measured abundances in C-rich AGB stars \citep{Agundez2017,Massalkhi2019}, indicating that this component does derive from acetylene-rich stellar material.

\section{Conclusions}
\label{section:conclusions}
\begin{figure}
\begin{center}
{\includegraphics[width=0.44\linewidth]{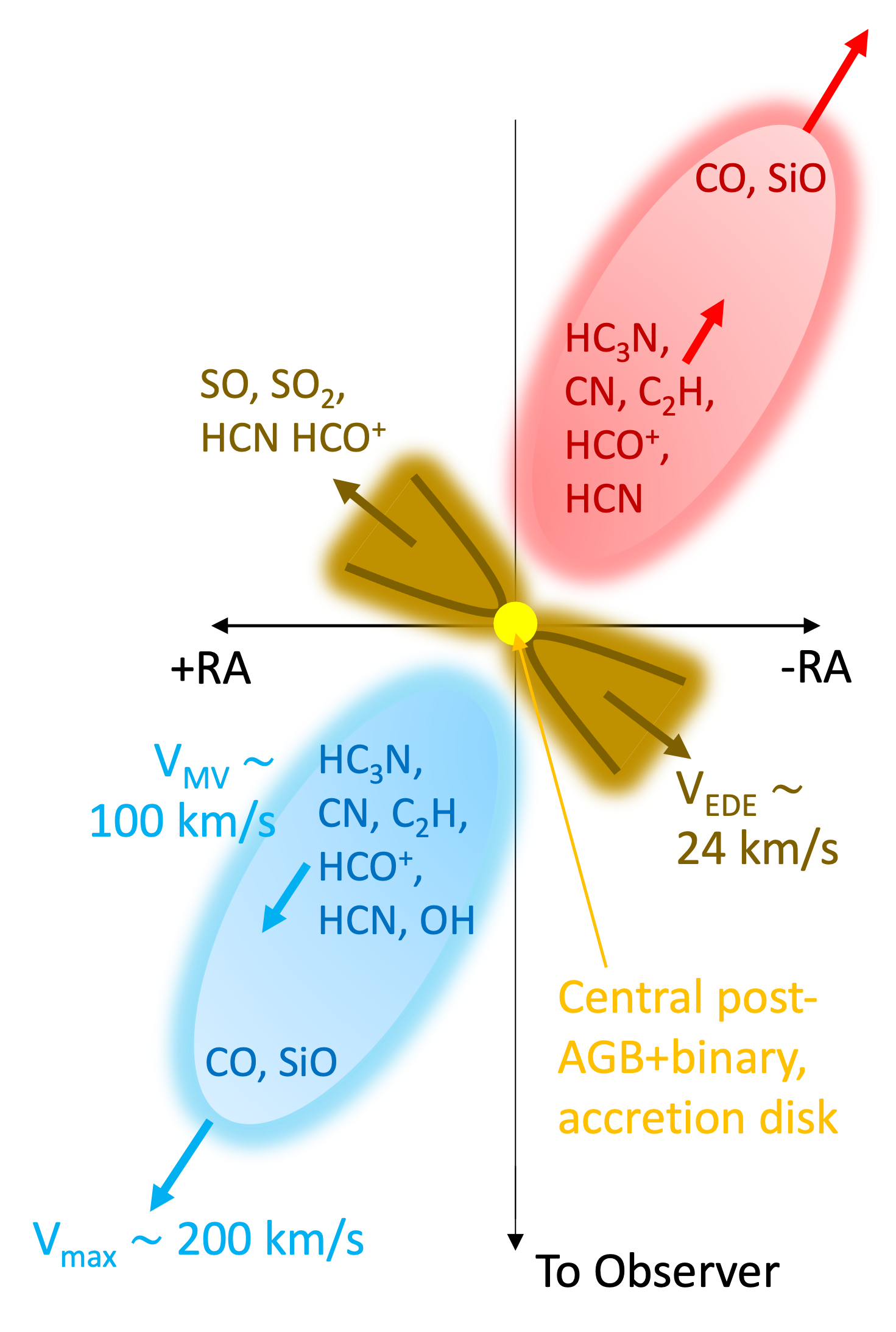}}
\caption{Diagram of the circumstellar environment of V510 Pup as suggested by our ACA analysis of its millimeter spectrum. Includes major morphological structures and their velocities, as well as the observed chemistry in each region. Proposed components include a carbon-rich bipolar outflow (red and blue), and a slowly expanding oxygen-rich equatorial waist surrounding the central binary (brown).}
\label{fig:schem}
\end{center}
\end{figure}

Using archival ACA observations, we analyzed the millimeter spectrum of V510 Pup to provide some of the first constraints on its molecular inventory. The new sensitivity and velocity information offered by our spectra indicate that not only does V510 Pup exhibit mixed chemistry in its outflow, but the carbon- and oxygen-rich material is separated into reservoirs of gas that are morphologically unique. The line profiles of the newly characterized CV component appear remarkably similar to the ``disk-prominent" subclass of post-AGB stars \citep{GallardoCava2022}, as well as the central region of the similar PPN HD 101584 \citep{Olofsson2019}. These equatorial density enhancements (EDEs) are common in PPNe and PNe, and recent works suggest that they can form during the AGB phase through binary-shaping \citep{Sahai2022,Decin2020}. Thus, one possible explanation for the observed chemistry (summarized in Fig.\ \ref{fig:schem}) is that V510 Pup harbors a slowly expanding waist from an earlier stage of evolution when its surface chemistry was O-rich, then as its fast bipolar outflow formed it also underwent a transition to ejecting C-rich material. This scenario has been proposed to explain mixed dust features in silicate carbon stars \citep{Ohnaka2013}, but the observations presented here provide some of the first kinematic evidence for it. To confirm this, future observations of V510 Pup at higher resolution will be critical, as spatially resolving the chemically distinct CV, MV, and HV components will allow one to determine their physical conditions and dynamical histories.

This paper makes use of the following ALMA data: ADS/JAO.ALMA\#2018.A.00047.S. ALMA is a partnership of ESO (representing its member states), NSF (USA) and NINS (Japan), together with NRC (Canada), MOST and ASIAA (Taiwan), and KASI (Republic of Korea), in cooperation with the Republic of Chile. The Joint ALMA Observatory is operated by ESO, AUI/NRAO and NAOJ. The National Radio Astronomy Observatory is a facility of the National Science Foundation operated under cooperative agreement by Associated Universities, Inc. Support for this work and travel was provided to M.\,A.\,S.\, by the NSF through the Grote Reber Fellowship Program administered by Associated Universities, Inc./National Radio Astronomy Observatory.

\end{document}